# Solar Activity, Solar Irradiance and Earth's Temperature[1]

**Professor Valentina Zharkova**

Summary

The Sun is the source of energy for the Earth and planets. The Sun has a magnetic field with north and south poles switching every 11 years and so does solar activity and radiation, which in the past modulated the terrestrial temperature. Solar activity is usually classified by numbers of sunspots, or magnetic loops, appearing on the solar surface generated by solar dynamo mechanism acting in the solar interior. The periodic occurrence of sunspots on the surface is found to be modulated by the solar background magnetic field and can be used as a new proxy of solar activity. Principal components analysis of solar magnetic field in cycles 21-23 detected pairs of magnetic waves generated by double dynamo in two layers of the solar interior with slightly different frequencies and a phase difference. These magnetic waves are reproduced with mathematical formulae enabling to discover of a grand solar cycle of 350–400 years, caused by the interference of these waves. When the waves are in anti-phase, the solar magnetic field and solar activity are significantly reduced indicating grand solar minima (GSMs). The previous GSM known as the Maunder minimum was recorded from 1645 to 1715 lasting for six cycles of eleven years. The modern GSM starts in 2020 and will last for three solar cycles until 2053. The solar radiation and terrestrial temperatures during this GSM are expected to be reduced but not as low as during Maunder minimum. This is because the modern GSM is much shorter; and because the terrestrial temperature has been increasing by the similar amount since Maunder minimum.

---

[1] This chapter is based on the original research published by *Nature Research* but subsequently retracted claiming concerns about the interpretation of how the Sun-Earth distance changes over time. Yet the original paper entitled 'Oscillations of the baseline of solar magnetic field and solar irradiance on a millennial timescale', did not actually include any new measurements of changes in this distance. Rather Zharkova et al. (2020) relied upon previously published work and an ephemeris for these calculations. What is unique about the now retracted paper is that it explains the importance of solar forcing caused by changes in the solar inertial motion (SIM) for accurate calculation of changes in global temperature. This phenomenon has so far not been included in any of the general circulation models relied upon by the Intergovernmental Panel on Climate Change (IPCC), and could explain much of the global warming over the last few centuries. Appendix 1 to this chapter provides a proof that the Sun-Earth distances in elliptic orbit, taken from the available ephemeris, are slowly changing in the past and current millennia because the Sun in its SIM shifts from the ellipse focus towards the spring equinox and aphelion of the Earth orbit. Appendix 2 presents a proof of significant variations of solar irradiance over the two millennia related to these changes of S-E distances exactly as it has been hinted at in the retracted paper.

The current study also uses the ephemeris of the Sun-Earth distances and associated variations of solar irradiance during the Earth revolution about the Sun. The ephemeris of the S-E distances show significant decreases of these distances by 0.005 au in 600-1600 by more than 0.01 au in 1600-2600 imposed by the solar inertial motion, or the deviation of the Sun–Earth distances from Kepler's third law. These Sun-Earth distance variations are reflected in oscillations of the baseline magnetic field reported by Zharkova et al, 2019. The S-E distance variations are followed by increases of solar irradiance in the first six months of each year in the two millennia 600-2600, which are not fully offset by the solar radiation decreases in the last six months. This study also reveals a continuous increase by more than 2.8% of solar irradiance input to the Earth atmosphere in the second millennium (1600-2600) compared to the first (600-2600). This solar radiation misbalance creates an annual surplus of the solar radiation in millennium 1600-2600 to be processed by the terrestrial atmosphere and ocean environments. Therefore, during the modern GSM (2020-2053) the two processes: decrease of solar activity and increase in total solar irradiance because of SIM, will result in a reduction of the current terrestrial temperature to levels reached in 1700, just after Maunder minimum.

**Figure 1.** Solar surface (photosphere) (top left), sunspots on the photosphere (middle left) as the footpoints of the embedded magnetic loops with opposite magnetic polarities (top middle and right). Change of magnetic polarities of the Sun during an eleven-year cycle (bottom plot).

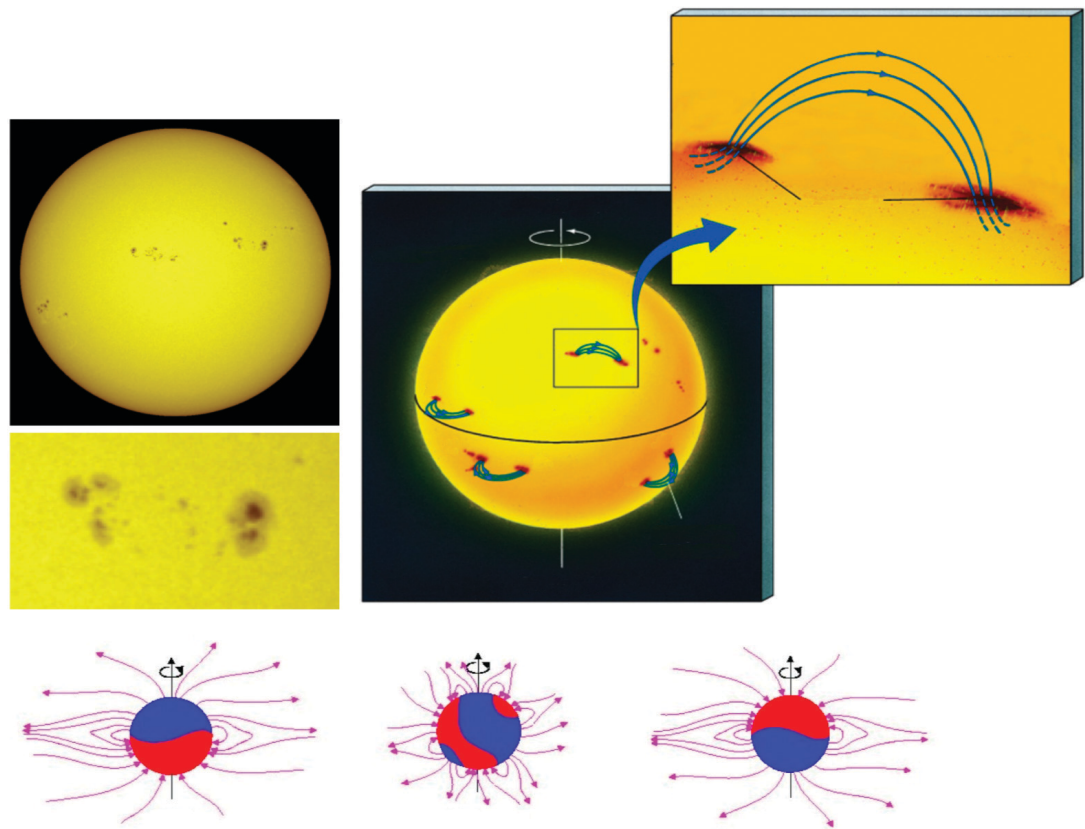

## 1. Sunspot cycles, solar irradiance and terrestrial temperature

Sunspots are dark features on the solar surface called photospheres (Figure 1, top left) appearing at start of each cycle at latitudes of about 30 degrees in the Northern and Southern Hemispheres (Figure 1, top middle and right) ) forming sunspot groups and migrating towards the equator as the cycle progresses (Spoerer & Maunder 1890). The polarities of leading and trailing sunspots in each hemisphere are maintained during a given cycle and change to the

**Figure 2.** Correlations between sunspot number (top plot), solar irradiance (middle plot) and terrestrial temperature (bottom plot).

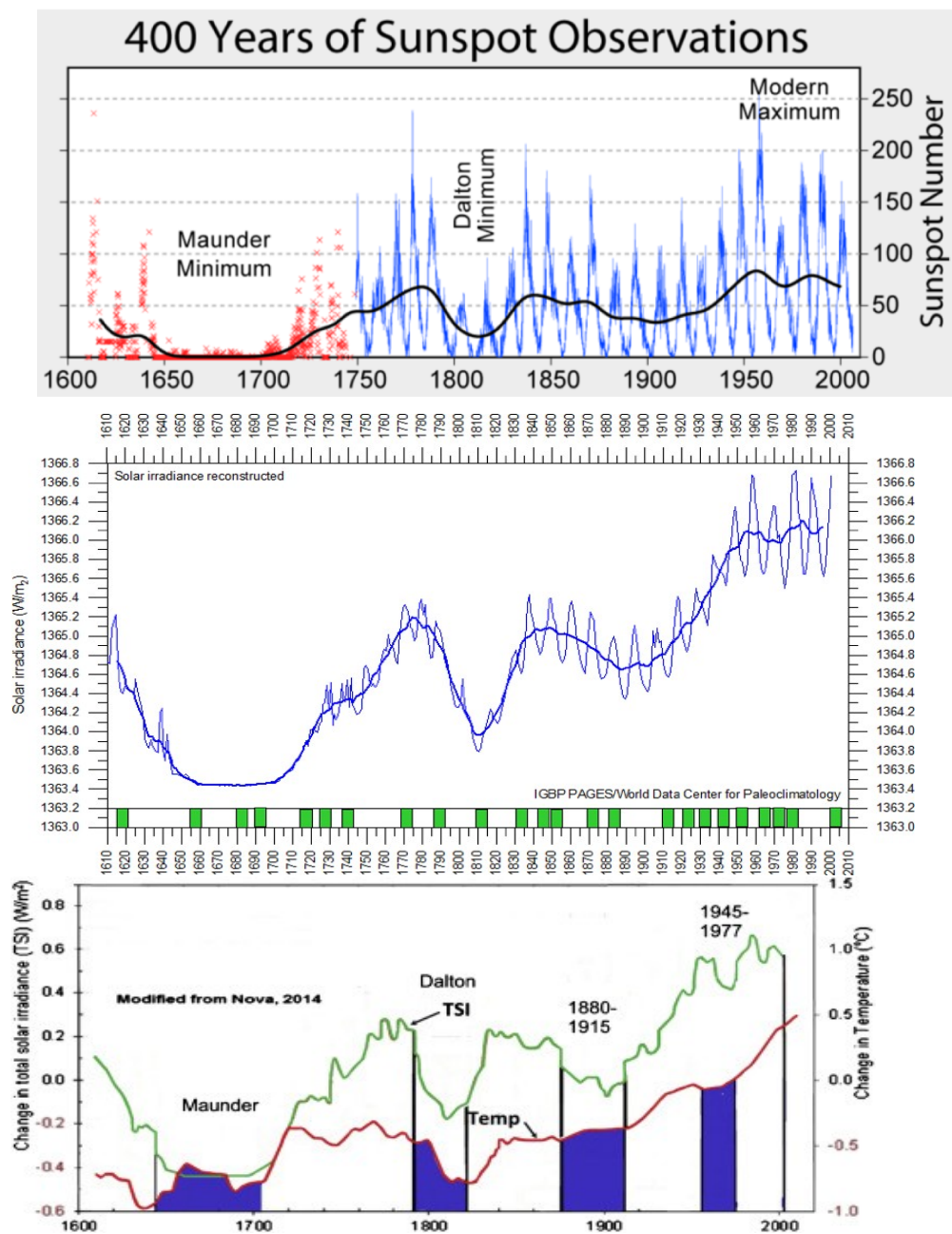

Source: Top: Belgian Royal Observatory; Middle: Lean et al. 1995; Bottom: Lee III et al. 1995; and Easterbrook 2016.

opposite polarities in the following cycle (Hale's law) (Hale et al.,1919) as shown in Figure 1 (bottom) with a change of north (blue) and south (red) magnetic polarities during an eleven-year cycle. Tilts of sunspot groups (SGs), or inclination of SGs towards the solar equator, are shown to grow with heliospheric latitude (Joy's law) allowing recycling of magnetic polarities to the poles for the next cycle.

An index of solar activity (Wolf 1870) is defined by the average numbers of sunspots and groups that occurred on the solar disk in a given month (Figure 2, top plot). Spoerer & Maunder 1890; Eddy 1976 established that there were periodic changes in the numbers of sunspots on the solar surface, which are considered to be variations in solar activity of the eleven-year cycle (Wolf 1870), or the 22-year cycle, when the full change of sunspot magnetic polarity is considered. The Sun is entering cycle 25, as recorded since 1755.

There were far fewer sunspots seen during some periods, for example, during the Dalton minimum (1790–1820), and practically none during the period known as the Maunder minimum (1645–1715). During the Maunder minimum, solar activity was significantly reduced for long periods of time (Figure 2, top) and so was the terrestrial temperature in the Northern hemisphere, as shown in Figure 2 (bottom plot). This was considered to be a result of a reduction of solar irradiance during the Maunder Minimum (Figure 2, middle plot).

Such dramatic reductions in solar activity, which are longer than a single eleven-year sunspot cycle, are known as grand solar minima (GSMs). Until recently, the physical mechanism causing these measurable reductions in solar activity beyond a single sunspot cycle were unknown, thus, making difficult to accurately predict such the extended periods of low solar activity.

## 2. Problems with predicting solar activity

The prediction of solar activity during a single solar cycle through averaged sunspot numbers has been used for decades for testing the accuracy of solar dynamo models, including the processes providing production, transport and disintegration of solar and sunspot magnetic fields in the solar interior and on the surface.

These cycles of magnetic activity are associated with the action of a dipole solar dynamo mechanism called $\alpha - \Omega$ dynamo (Parker 1955; Babcock 1961; Leighton 1969). It assumes the solar dynamo occurs in a single spherical shell of the solar interior at the bottom of solar convective zone (SCZ), or tachocline, where twisting of the magnetic field lines ($\alpha$-effect) and the magnetic field line stretch and wrap around different parts of the Sun ($\Omega$ −effect) acting together as the solar dynamo mechanisms (Brandenburg & Subramanian 2005; Jones et al. 2010).

This dynamo theory of generation of toroidal fields of sunspots by dipole magnetic sources from the background poloidal field defining solar activity has significantly progressed in the past three decades (see, for example, Cameron et al. 2016, 2017).

Solar activity in cycle 24, however, was much lower than in the previous three cycles 21–23 and as predicted by many models (Pesnell, 2008) besides a few model-based predictions considering polar magnetic fields of the Sun (e.g. Choudhury et al. 2007). The reduction in solar activity in cycle 24 was surprising because the previous five cycles were extremely active, forming the Modern Maximum (Usoskin et al. 2016). These persistent disagreements between the measured and predicted sunspot numbers were indications of some missing processes defining solar activity by sunspot numbers.

Then the best pattern-recognition methods were applied to fully automated detection of sunspot and their magnetic fields (Zharkov et al. 2005; Zharkova et al. 2005) that confirmed Hale's and Joy's laws (Zharkov et al 2006; Zharkova & Zharkov 2008). However, the automated detection did not significantly improve prediction of solar activity allowing scientists to reliably predict a solar cycle only after it had already started (Karak & Nandy 2012).

## 3. Role of the solar background magnetic field

It was found (Babcock, 1961) that weaker background magnetic field surrounding sunspots has polarities opposite to the leading sunspot polarities. The background magnetic field polarities in the opposite hemispheres were consistent with the action of a magnetic dipole, whose sign is changed in line with the change of sunspot magnetic field polarities. By comparing solar background and sunspot magnetic fields for cycle 23, Zharkov et al. 2008 found the magnetic polarities of background magnetic field and leading polarities of sunspots, to be in anti-phase. Furthermore, the solar background magnetic field (SBMF) was found to be the leading force defining timing and locations of sunspot occurrences and migration on the solar surface, accounting for their biennial cycle of 2.5 years and the north–south asymmetry of sunspot areas and magnetic field magnitudes (Zharkov et al. 2008).

This highlighted a leading role of SBMF in solar activity and inspired us to apply principal component analysis (PCA) to solar background magnetic field (Zharkova et al. 2012). PCA acted on SBMF as a glass prism does on white light, obtaining a rainbow of light at different wavelengths from red to ultraviolet (Zharkova et al. 2018a). This approach allowed us to split complex magnetic waves seen on the solar surface, or the photosphere, into separate components, leading to reliable description of solar activity with mathematical formulae giving possibility of its forecasting on a long timescale.

By applying principal component analysis (PCA), four pairs of waves were identified on the solar surface. The first pair of principal components (PCs) defines the strongest waves in solar background magnetic field, covering about 67% of the data (by standard deviation) (Shepherd et al. 2014; Zharkova et al. 2015). These two PCs represent magnetic waves with slightly different frequencies and an increasing phase shift (see Figure 3, top). These waves start in the opposite hemispheres while traveling to Northern hemisphere in odd cycles and to Southern hemisphere in even cycles (Zharkova et al. 2012, 2015).

These two magnetic waves are found generated by a double electro-magnetic dynamo from the dipole magnetic sources acting in two layers of the solar interior: inner (bottom of SCZ) and shallow (under the surface) outer layers, each of which has different meridional circulation velocities (Zharkova et al. 2015). The existence of such two layers in the solar interior was confirmed by helioseismological observations using a Helioseismic and Magnetic Imager (HMI) aboard the Solar Dynamic Observatory (Zhao et al. 2013).

A summary curve derived from the two PCs indicated the overall variations of magnitude and polarity of the pair of strongest magnetic waves (see Fig. 3, bottom plot). Each PC was quantified with the formulae containing series of five periodic functions derived using symbolic regression analysis (Shepherd et al. 2014; Zharkova et al. 2015). These observational PCs are shown to be well defined by the model dynamo waves produced by dipole magnetic sources in two (inner and outer) layers of the solar interior with slightly different velocities of meridional circulation that imposes slightly different frequencies of waves (Zharkova et al. 2015).

The summary curve of these two PCs provides the SBMF polarity as well as the amplitude of magnetic field in a current cycle. This summary curve was found to decrease in the past 3 cycles (21-23) and continue decreasing for the next three cycles 24-26 (Figure 3, bottom plot). The modulus of the summary curve is found to fit rather closely the existing solar activity curve in cycles 21–23 defined by average sunspot numbers (Fig. 4, top plot) even uncovering the error in its definition in the second half of cycle 23 also reported by Clette et al. 2015. Hence, the summary curve was suggested as a new proxy of solar activity because it defines not only an amplitude but also a magnetic field polarity.

If maxima of the two waves occur in the same hemisphere and their phases are rather close, the two waves have constructive (resonant) interference leading to a maximum of the summary, or solar activity, curve. For waves with the phase shift large enough, a double maximum of the summary curve naturally occurs (Gnevyshev & Ohl 1948). The hemisphere, where the constructive interference happens, becomes more active one, naturally accounting for the north–south asymmetry of solar activity in cycle 23 (Zharkov et al. 2005; 2008) and in a few other cycles (Temmer et al. 2002; Belucz & Dikpati 2013; Shetye et al. 2015).

The mathematical formulae were derived for temporal evolution of the two PCs, allowing us to calculate the summary curve, or solar activity index, either forwards or backwards in time. In Figures 4 (bottom plot) we present the modulus summary curves calculated for cycles 24–26 (Shepherd et al. 2014), which, first, fits well the solar activity index for cycle 24 and, second, reveals a significant decrease of solar activity in cycle 25 and, especially, in cycle 26. The summary curve of PCs proves a powerful proxy of solar activity, which also considers a magnetic polarity of solar background magnetic field, or the leading polarity of sunspots. This approach enables reliable prediction of solar activity on longer time scales based on the derived summary curve, which is shown to be closely linked with the double dynamo model induced by dipole (Zharkova et al. 2015) or dipole plus quadruple (Popova et al. 2018) magnetic sources in the inner and outer layers of the solar interior.

In order to test the further predictions of solar activity with the summary curve, let us extend this curve from the current time (2015) forwards to 3200 and backwards to 1200 (see Fig. 5, top plot). This allowed us to discover the regular grand solar cycles (GSCs) of solar activity with a duration of 350–400 years, evidently caused by the interference (beating effect) of the two magnetic waves with close but not equal frequencies generated by the double dynamo action in these two layers of the solar interior (Zharkova et al. 2015). These grand solar cycles are separated by grand solar minima (GSMs) marking the years between the GSCs with nearly zero solar background magnetic field, or, consequently, with negligible solar activity, similar to that predicted in cycles 25–27.

**Figure 3**. The two PCs and a summary curve of the SBMF

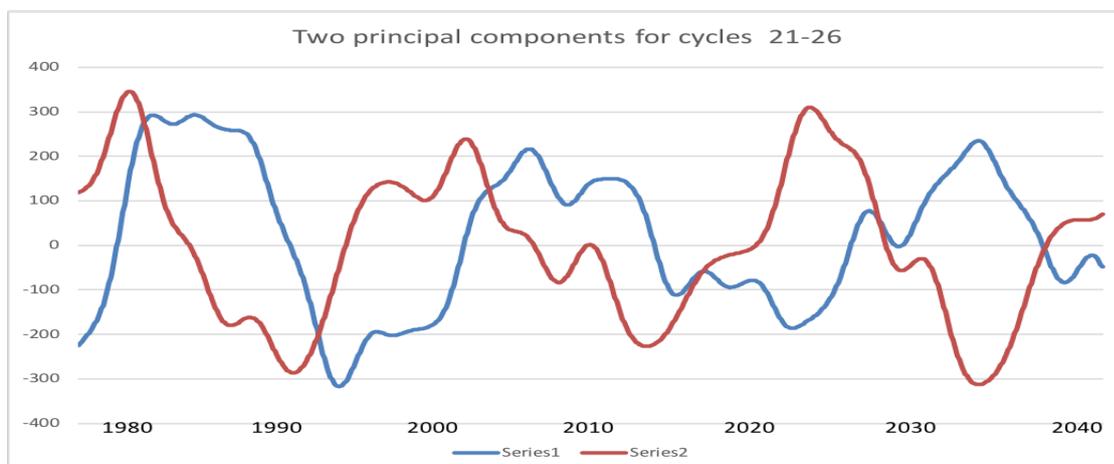

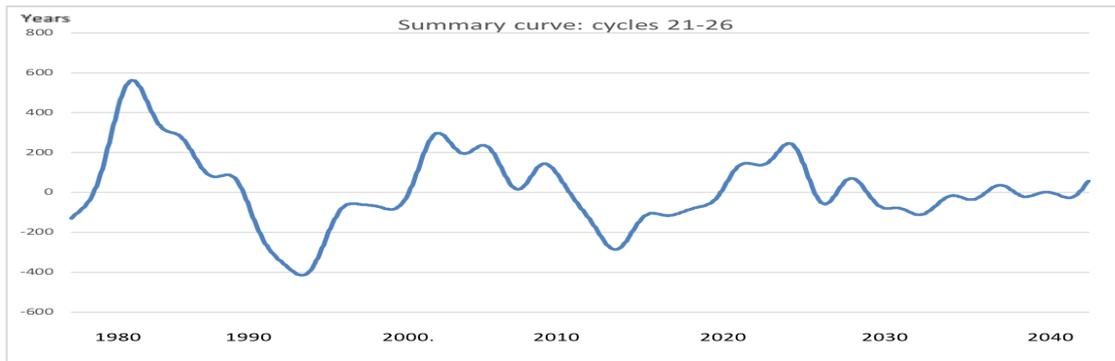

Top: The two PCs of SBMF obtained for cycles 21–23 using historical data and predicted for cycles 24–26.

*Source: Zharkova et al. 2012; Shepherd et al. 2014; Zharkova et al. 2015*

Bottom: The summary curve of two PCs indicating magnetic field amplitudes and polarities derived from these two PCs for the 'historical' (cycles 21–23) and predicted for cycles 24–26 showing a reducing solar activity.

*Source: Zharkova et al. 2015*

The timings of previous GSMs shown in the summary curve of Figure 5 (top plot) closely fit the Maunder minimum (1645–1715) and Wolf minimum (1280–1350), and predict the two upcoming modern GSMs (2020–2053 and 2375–2415). Furthermore, by extrapolating the summary curve backwards 3000 years to 1000 BC (see Fig. 5, bottom plot, blue-coloured curve) (Zharkova et al. 2018b) it reveals the further GSMs – Oort's (1040–1080), Homeric (550-400). This clearly gives a better accuracy of solar activity definition, in comparison with the prediction of sunspot activity restored from the past TSI derived with carbon-14 dating (Solanki et al. 2004).

Figure 4. The modulus summary curve as a new proxy of solar activity

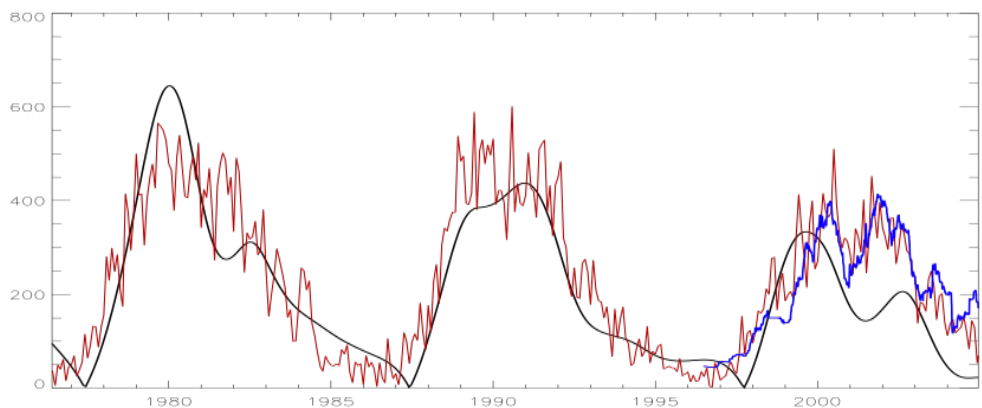

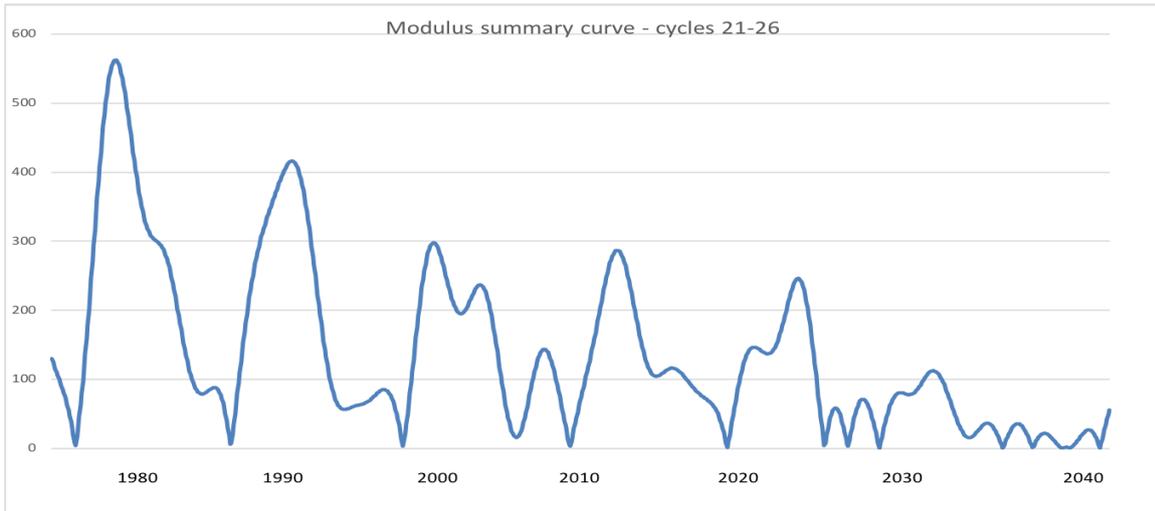

Top: The fit of modulus summary curve to average sunspot numbers for cycles 21–23.

Bottom: The modulus summary (solar activity) curve derived for cycles 21-23 and that predicted for cycles 24–26. *Source: Shepherd et al. 2014; Zharkova et al. 2015*

**Figure 5.** Solar activity prediction forward and backward over millennia using the summary curve of PCs.

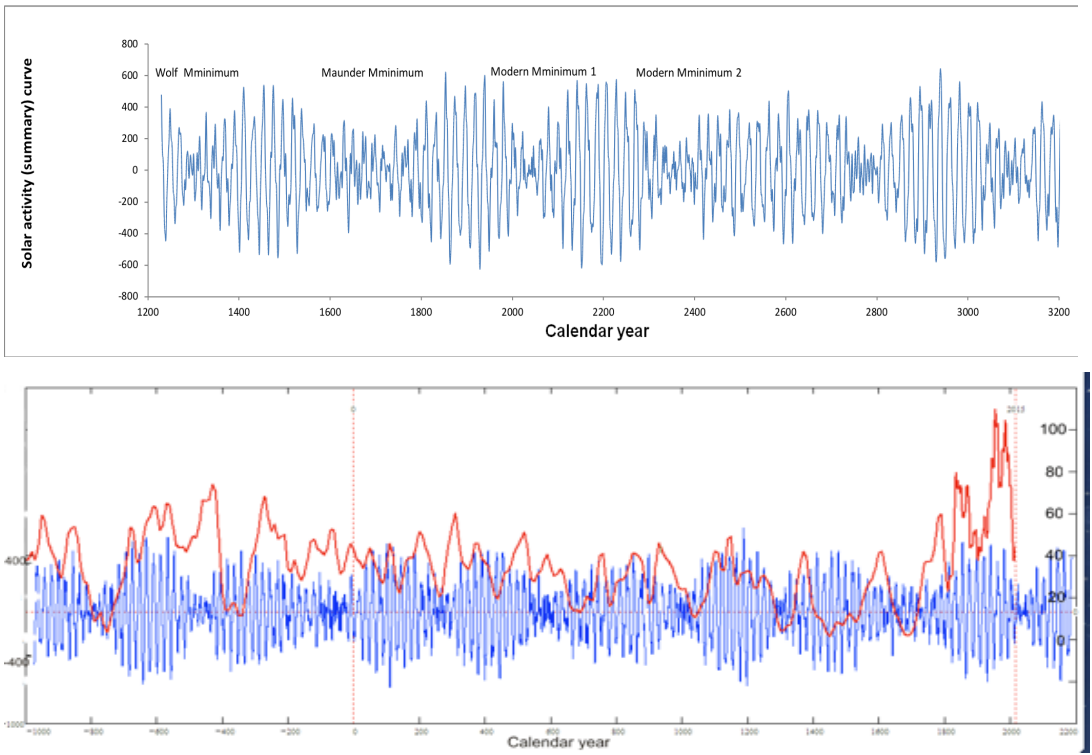

Top: the summary curve (arbitrary units, Y-axis) from two PCs extended in years (X-axis) forward to 3200 and backward to 1200 (Zharkova et al. 2015).

Bottom: the summary curve (blue line, arbitrary units) extended backwards by 3000 years to 1000 BC versus the average sunspot numbers (red line) restored for the same period from the carbon 14 dating (Solanki et al. 2004).

*Source: Top: Zharkova et al. 2015; Bottom: Zharkova et al. 2018b*

## 4. Super-grand Hallstatt's cycles

By calculating the residuals of large-scale magnetic field oscillations of the redacted summary curve restored back (hindcasting) to 120,000 years, and by deriving magnitudes of the baseline (zero-line) magnetic field for each 22-year set, we discovered very rigid periodic variations of the baseline magnetic field, shown in Figure 6 (top plot) by the dark-turquoise line over-plotted onto the summary curve.

**Figure 6.** Oscillations of the baseline magnetic field with a period of about 2000 - 2100 years

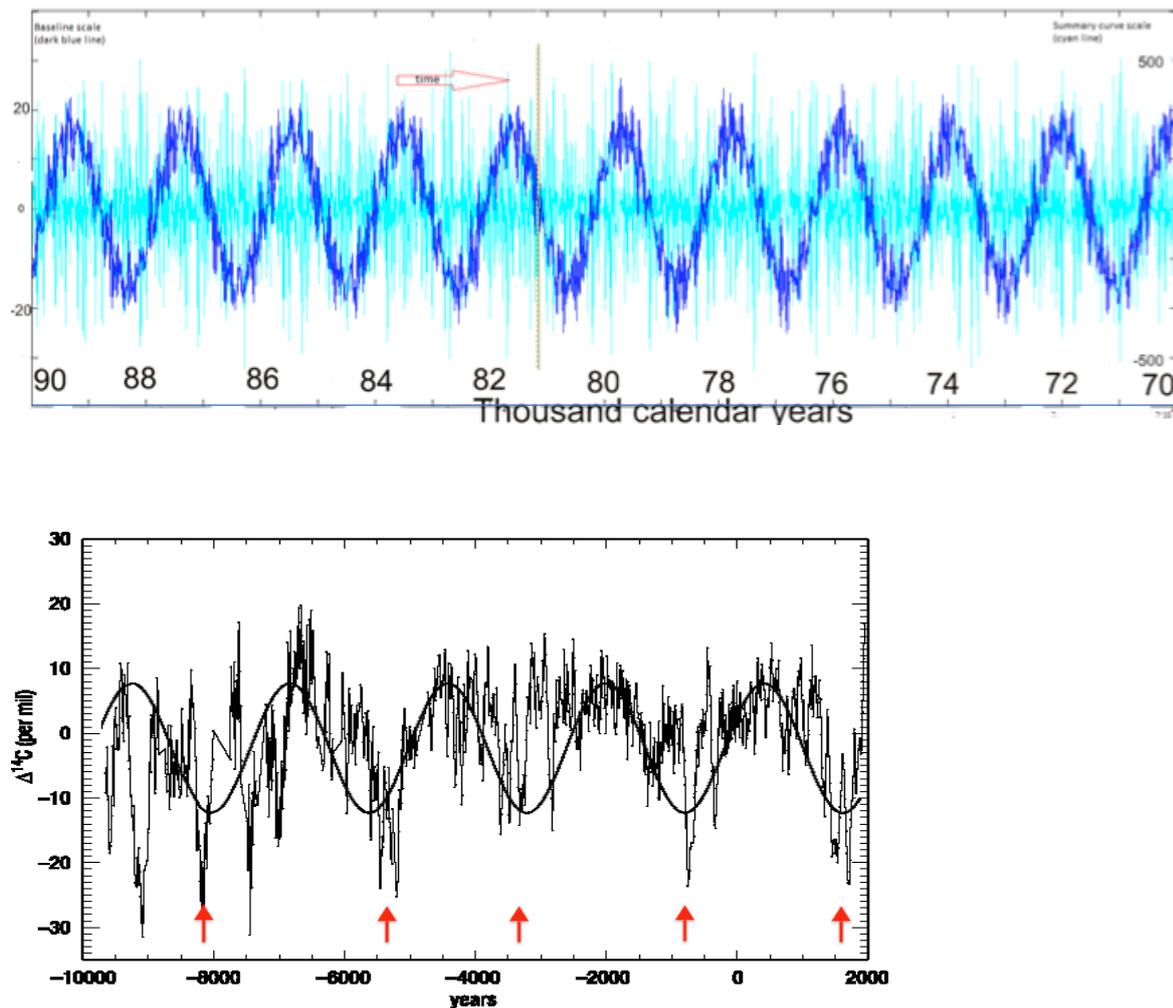

Top plot: Periodic variations of the baseline magnetic field (super-grand cycle of 2100 years) shown by the dark turquoise curve (left Y axis) versus the redacted summary curve (bright blue curve, right Y axis). The baseline variations are those of a zero-line of the summary curve, which are too small to observe on this curve without filtering out the large-scale 22- year oscillations of magnetic field amplitudes.

Source: Zharkova et al. 2019; 2020

Bottom plot: Oscillations of carbon-14 abundances during the Holocene (Reimer et al. 2009). The purple arrows indicate the recorded minima of carbon-14 abundances, which are close to the minima of the super-grand Hallstatt's cycles derived from the summary curve in the top plot (Zharkova et al. 2019).

Source: Reimer et al. 2009

The baseline magnetic field oscillations show a very stable period of about 2000–2100 years through the past 120,000 years that is very similar in length to the period of 2200 years of Hallstatt's cycle reported from restoration of solar irradiance in the Holocene (Vieira et al. 2011; Steinhilber et al. 2009; Steinhilber et al. 2012). The two-millennial oscillations of the magnetic baseline are followed by the similar oscillations of carbon-14 abundances, derived in the tree rings (Reimer et al. 2009) shown in Figure 6 (bottom plot). The carbon-14 isotope, also known as radiocarbon, is produced in the upper atmosphere as a biproduct of cosmic-ray collisions or of collisions with energetic protons coming from solar flares and coronal mass ejections.

The oscillations of total solar irradiance (TSI) during the Holocene (Steinhilber et al. 2009; Vieira et al. 2011) are presented in Figure 7 (top plot) showing TSI to have a deep minimum near the Maunder minimum and to increase towards modern times. For the modern super-grand cycle shown in Figure 7 (bottom plot), the TSI (orange line) is found to follow closely a slope of the baseline magnetic field curve (dark blue line) and the slope of terrestrial temperature (black line). From a current growth of the baseline magnetic field curve one can anticipate the solar irradiance curve (Steinhilber et al. 2009; 2012) to continue growing until its maximum in 2600, as shown in Figure 7 (bottom plot).

**Figure 7.** Modern super-grand cycle and TSI variations

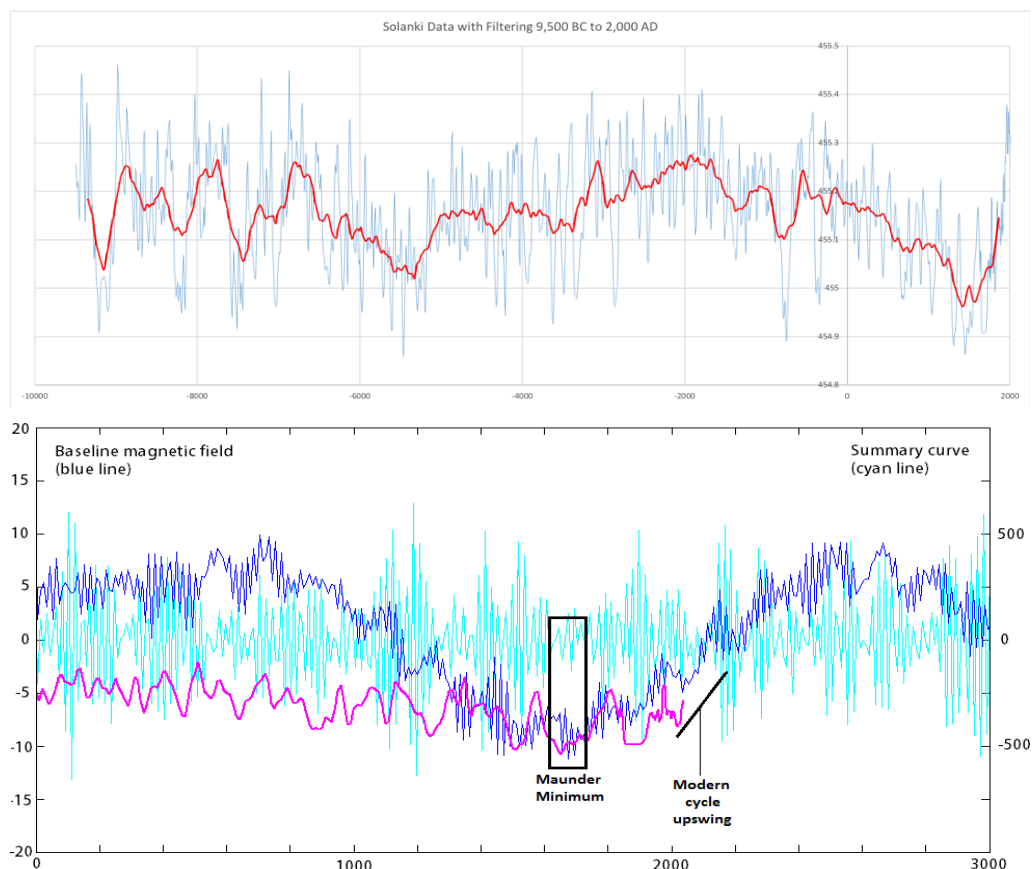

Top plot: Solar irradiance curve for the Holocene, the last 12,000 years.

*Source: Modifed from Steinhilber et al. 2009; Vieira et al. 2011.*

Bottom plot: The baseline magnetic field cycle (dark turquoise), TSI (orange line), redacted summary curve of magnetic field (blue line) showing the super-grand period of the modern cycle upswing. The black line shows the slope of the baseline terrestrial temperature increase derived from Akasofu (2010) (Figure 10.9, bottom plot). Note, the orange line of TSI curve was slightly reduced in magnitude, in order not to fully overlap with the magnetic field baseline oscillations.

Sources: Vieira et al. 2011; Zharkova et al. 2019; 2020; Akasofu 2010

## 5. Solar inertial motion and Sun–Earth distance variations

In order to understand the nature of the super-grand millennial (Hallstatt's) cycle, one needs to combine Kepler's laws for planetary orbits with the gravitational forces of the planets (mainly by Jupiter, Neptune, Saturn and Uranus) on the Sun and these orbits, called the solar inertial motion (SIM) (Jose, 1965; Fairbridge et al. 1987; Charvatova 1988, 2000; Palus et al. 2007).

Owing to SIM, the planets in the solar system are forced to move around about the centre of mass (barycentre) of the solar system (Bretagnon and Francou, 1988; Lascar et al, 2011, Folkner et al, 2014). SIM also makes the Sun to move in epitrochoid-shaped orbits (with a diameter up to 4.3 times its radius) shown in Figure 8 (top plot). This 'wobbling star' effect is

**FIGURE 8.** Solar wobbling in solar inertial motion (SIM).

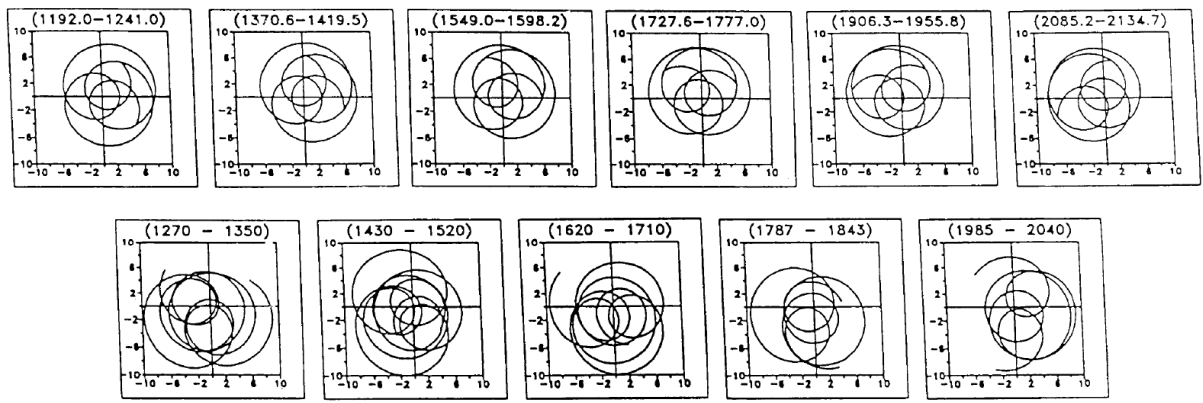

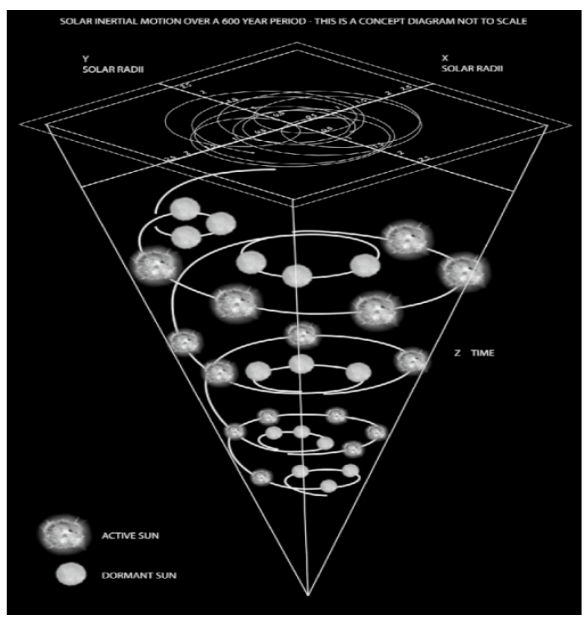

<u>Top:</u> The wobbling Sun orbits and the increasing-in-time radius of the SIM orbits.

<u>Bottom:</u>  the increasing SIM radius relative to the centre of mass (barycentre) of the solar system.

*Source: Adapted from Charvatova 1988; Palus et al. 2007; Mackey 2007.*

widely used in Astrophysics to detect the stars with planetary systems and exoplanets.

The radius of SIM orbits is varying in time as shown in Figure 8 (bottom plot) depending of the planet positions; thus, shifting the Sun location from the ellipse focus towards different parts of the planetary orbits (see Appendix 1 for the ephemeris of Sun-Earth distances).

**5.1. The Earth orbit.**  By exploring the JPL ephemeris for the Earth orbit https://ssd.jpl.nasa.gov/horizons.cgi#top we observed a small increase of the aphelion distance and similar decrease of the perihelion distance in the millennium from 1600 to 2600. However, these variations do not change much the Sun-Earth distances averaged by either of the orbit parameters (angle, time or arch) (Stein & Elsner, 1977). The variations of the aphelion and perihelion distances produce the reduction of eccentricity from 0.0170 in 1600 to 0.0163 in 2600. This would lead to a change of the average distance by time from 1.0001462 au in 1600 to 1.0001328 au in 2600, e.g. the difference is negligible. This indicates that the Earth orbit remains pretty stable elliptic one and does not fully follow the Sun in its SIM as many climate researchers think.

**5.2. The Sun-Earth distances from ephemeris.** Then we explored the daily Sun-Earth distances over the two millennia (600-2600) derived from the VSOP87 ephemeris, which coincide up to 6 decimal digits after the decimal coma with the JPL ephemeris.  The Sun-Earth distances are presented in Appendix 1 for the millenniums M1 (600-1600) (Figure A1.1) and M2 (1600-2600 (Figures A1.2) with annual S-E variations summarised in Figure 9 of this chapter. The Figures show in both millennia the clear shifts of the Sun's locations from the ellipse focus, where it is supposed to reside, according to Kepler's laws, towards the spring equinox of the Earth orbit caused by SIM.

This shift leads to a reduction of Sun-Earth distances (Figure A1.3) in January - June  up to 0.005 au (in April-May) per millennium M1 and up to 0.01 au per millennium M2 (in April-May) as expected from SIM ( Charvatova, 1988, Palus et al, 2007 and to increase of the S-E distances in August-December. The decrease of the S-E distances in the two millennia agrees with the recent calculations of Jupiter and Saturn effects on the Sun inertial motion revealing the oscillations with period of 4.3-4.4 thousand years (Perminov and Kuznetsov, 2018). This can explain the S-E distance decrease in January-June and increase in July- December during both millennia: smaller for  M1 and larger for M2 shown in Figures A1.1 and A1.2, which will be followed by the opposite trend in the next two millennia, when the Sun will relocate in its SIM towards the autumn equinox.

Although, Figure 9 demonstrates that these S-E distance decreases and increases in M1 (600-1600) (left plot) are nearly symmetric with respect to the winter and summer solstices, so their perihelion and aphelion distances are close to those of ellipse. While in M2 (1600-2600) these perihelion and aphelion distances are shifted by 20-25 days forward from the winter and summer solstices (to mid-January and mid-July, respectively), thus, creating the local perihelion and aphelion distances for a given year, which are slightly different from those defined for the ellipse with a star in its focus. This is demonstrated in more details in Appendix 1 (Figures A1.4 and A1.5) showing the S-E distance differences (monthly and annual, respectively) between the increases/decreases of the S-E distances in M1 and M2. Then the double differences were calculated by subtracting the difference for M2 from that of M1 (M1-M2) (Figure A1.5). It is evident that the double differences become negative in April and remain such until end of October meaning the S-E distance decrease in April -July and its increase in July-December in M2 (1600-2600) is ahead of that in M1 (600-1600) meaning there should be more solar radiation input to the Earth in millennium M2 as we show in the next section.

Moreover, these variations of the S-E distances shown in Figure 9 and Appendix 1 can explain the oscillations of the baseline solar magnetic field (Hallstatt's cycle) shown in Figure 7 (bottom plot) (Zharkova et al, 2019). In M2 the Sun moves in its SIM inside the Earth orbit on wider orbits shifting towards the spring equinox and summer solstice in the Northern hemisphere. Hence, in M1 the Sun location is much closer to the ellipse focus of the Earth orbit resulting in smaller magnitude of the baseline magnetic field of northern polarity, or minimum on the dark blue line in Figure 7. While in M2 the Sun shifts much further from the focus towards the spring equinox position of the Earth orbit, so that there is a shift of the S-E longest distance from 21 June (when the ellipse aphelion is approached) to mid-July.

This means during M2 the Earth stays a bit (20-25 days) longer being closer to the Sun before it approaches the local aphelion while it is turned towards the Sun with its North pole. This is causing a small rise to the northern baseline magnetic field shown in Figure 7 (dark blue line) lasting until 2600 (Zharkova et al, 2019, 2020). Therefore, this confirms the hint expressed in paper by Zharkova et al, 2019 explaining that the baseline magnetic field oscillations derived purely from magnetic field observations are, indeed, caused by the gravitational effects on the Sun of planets, or solar inertial motion.

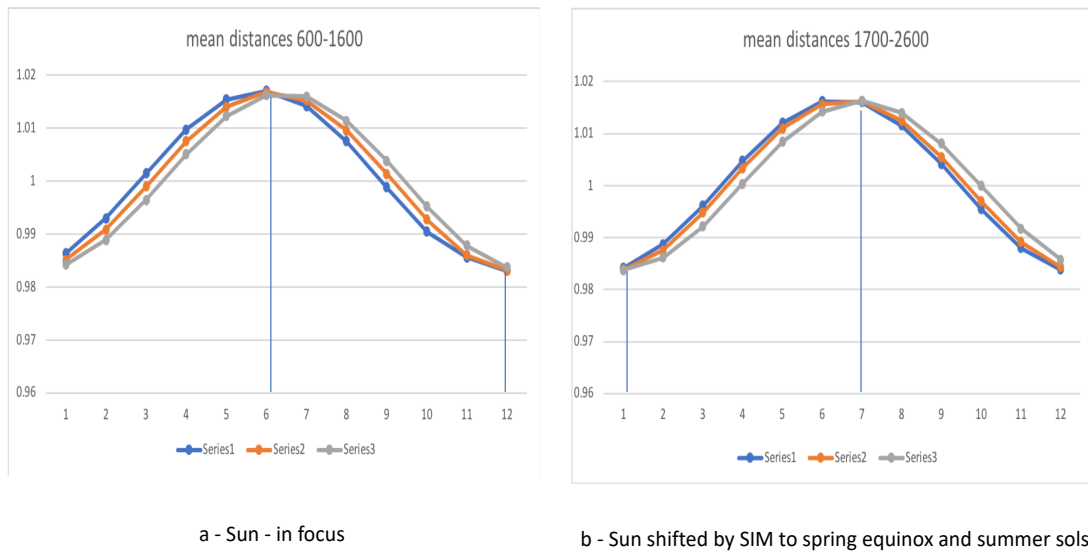

a - Sun - in focus  b - Sun shifted by SIM to spring equinox and summer solstice

**FIGURE 9.** Left: Annual variations of the mean monthly Sun -Earth distances (au) for the years of 600-1600.  Right: Annual variations of the mean monthly Sun -Earth distances (au) for the years 1700-2600 (right). X axis indicate months of a year. The vertical lines indicate times when the maximum distances (local aphelion) are achieved. Note, we selected for the second plot year 1700 in the right plot to avoid a duplication of the curve from the left plot.

## 6. Changes in total solar irradiance (TSI)

Changes in solar radiation amount reaching the Earth are affected by the changes:

- of distance between the Earth and the Sun in Hallstatt's cycles;

- of solar activity during grand solar minima (every 350 to 400 years), 3 W/m², or 0.22% (Lamb 1972; Lean et al. 1995; Fligge & Solanki 2000);

- of solar activity during eleven-year sunspot cycle, about 1.3 W/m², or 0.1% of TSI (Wilson et al. 1991; Lee III et al. 1995).

### *6.1. TSI variations with distance*

Following variations of the S–E distances (Appendix 1) let us evaluate variations of TSI  in the millennia M1 and M2 using the method of inverse squares discussed in Appendix 2.  For the TSI normalisation the magnitude of 1366 W/m² (Lean et al, 1995) for  the longest distance in June 1700 is used. The TSI daily increases for every month are presented in Figure A2.1 (for M1) and Figure A2.2 (for M2) with their annual variations compared in Figure A2.3.

The increase of solar irradiance during the months January-June in M1 is nearly balanced by its decrease from July to December (Figure A2.1 in Appendix 10.2) while in M2 this TSI curve is shifted to February-July for the atmospheric heating and August- January for atmospheric cooling (Figure A2.2). The annual variations of the monthly averaged TSI magnitudes (Figure A2.3) reveal a steady increase of solar irradiance during spring-summers and decrease during autumn-winters in the Northern hemisphere in each century *occurring on the top of regular seasonal variations* because of the S-E distance variations shown in Figure 9. There is also a shift of the minimum point of the TSI annual variations (Figure A2.3) from 21 June (in M1) to mid-July (M2) that indicates possible disbalance between the annual TSI input and output in M2 (1600-2600).

Because of the reduction of a S–E distance caused by SIM, the TSI is increased from 1700 to 2600 by about 13 W/m$^2$ (0.95%) in February-March (and increased by the same amount in August-September), by 18 W/m$^2$ (1.3%) in April-May (decreased in October-November) and by 7 W/m$^2$ (0.5%) in June-July (decreased in December-January) (see Figures A2.1-2.2). These numbers can be added to produce more than 2.8% of solar irradiance increase that is comparable with the estimations up to 3.5% hinted in the retracted paper by Zharkova et al., 2019.

Based on the location of Earth on orbit, this TSI inputs are divided unevenly between the hemispheres depending, which one of them is turned towards the Sun. This means that, because of the Earth tilt of 23.5º from the vertical to the ecliptics in M2 (1600-2600), the Northern hemisphere will have an increasing input of solar radiation during the months June – July, while in the previous millennium M1 (600-1600) the input of solar radiation was reducing from 21 June and keeps reducing through the whole July. This means that a decrease of the solar input in November-December-January must be lagging its increase in April-July, since according to the Kepler's second law, the Earth moves quicker at the part of the orbit in June-July than in December-January, thus, passing quicker the positions with a reduced radiation in December-January than with the increased one in June-July.

These variations explain a wide variety of the measured TSI magnitudes in the earlier space observations of 1370 W/m$^2$ (Shirley et al. 1990), 1971 W/m$^2$ (Wolff & Hickey 1987), or 1972 W/m$^2$ (Lee III et al. 1995). The annual variations of the mean TSI for every month plotted in Figure A2.3 for M1 (left plot) and M2 (right plot), show that the minimum of the mean TSI curve shifts towards the mid-July, thus, securing extra heating of Northern atmospheres in the summer months (second half of June and firth third of July). These shifts of the largest S-E distances from 21 June to 15 July in M2 can also explain why the baseline solar magnetic field is an ascending phase of Hallstatt's current cycle, with a maximum of the northern polarity at 2600 before the longest distance (local aphelion) becomes shifting back to the ellipse aphelion in June in the next millennium. As shown above (Figure 6, bottom plot) there have been about 60 super-grand Hallstatt's cycles over the past 120,000 years (Zharkova et al. 2019; 2020) meaning such the millennial changes of TSI on Earth are regular patterns, which will continue in the current Hallstatt's cycle shown in Figure 7 (bottom plot).

**6.2. Disbalance of the TSI depositions in two millennia.** Having the daily magnitudes of TSI shown in Figures A2.1 and A2.2, it is possible to count the total annual amount of TSI emitted by the Sun towards the Earth in each year of the both centuries. If this amount does not change from year to year, then TSI is, indeed, the same for each year for both centuries, as currently assumed. However, the real annual TSI variations deposited to Earth are shown in Figure 10 for the two cases: a) the averaged monthly TSI magnitudes calculated for the S-E distances shown in Figure 9 when only 12 magnitudes per year (for 12 months) are added; b) the daily TSI magnitudes taken from Figures A2.1 and A2.2 associated with daily magnitudes of TSI (for 366 days for leap years).

It is evident that these two plots clearly show that owing to SIM the monthly TSI variations (case a) show the increase of TSI by about 1-1.2 W/m$^2$ in 2020 compared to 1700 (Figure 10, left plot). This TSI increase is close to a magnitude of 1-1.5 W/m$^2$ reported from the current TSI observations (Krivova et al., 2011). However, the annual TSI magnitudes, calculated from the daily S-E distances and solar irradiance data, reveal much larger annual increase of solar irradiance by about 20-25 W/m$^2$ (1.8%) in M2 by 2500 than in millennium M1 (Figure 10, right plot).

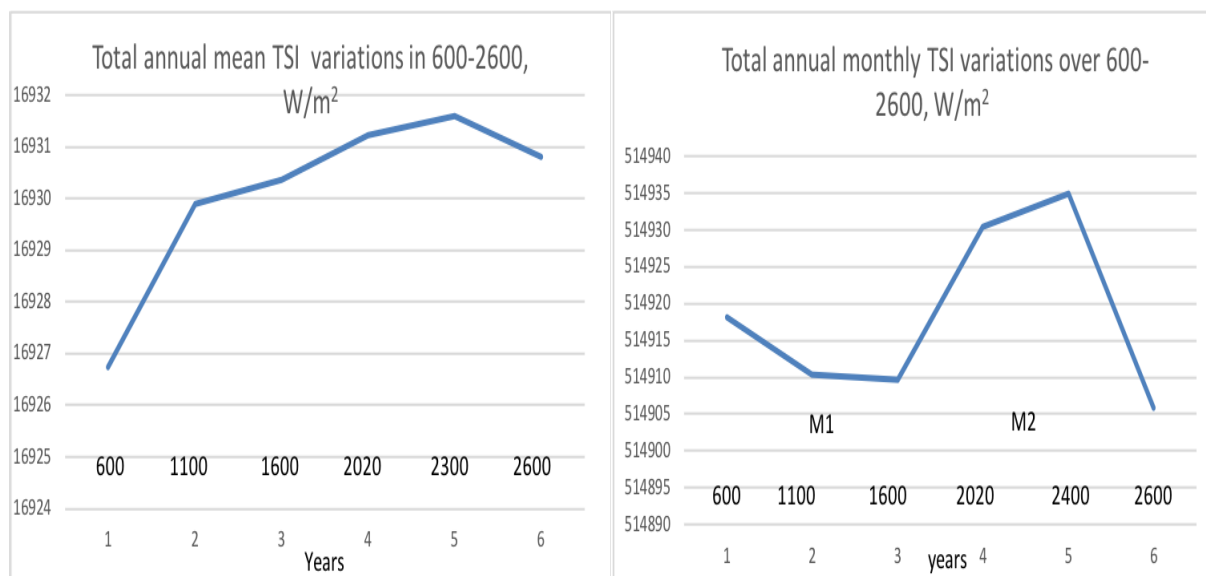

**Figure 10.** Millennial variations of the annual solar irradiance in the millennium M1: 600-1600 and M2: 1600-2600 calculated from the TSI averaged per month (left plot) and not-averaged (daily) TSI (right plot).

This analysis gives the indication of the averaged TSI increase per century by 2-2.5W/m², or (0.15-0.18)%, comparing to the TSI in 1700. This is the very important hidden solar irradiance input in millennium M2 (1600-2600) caused by the SIM effect, which was significantly underestimated if the averaged monthly TSI magnitudes are used (Figure 10, left plot).

The essential issue is how much of this extra solar radiation is distributed between the hemispheres, since it will not be even owing to the Earth tilt and its position on the orbit. At the start of the year, in January, the Earth is turned to the Sun with the southern pole, meaning that any decrease and increase of solar radiation is mostly absorbed by the parts in Southern hemisphere. Then when Earth orbiting approaching March, the distribution of solar irradiance between the hemisphere becomes nearly even, while in April-June the main part of the solar radiation is shifted towards the Northern hemisphere, having its maximum/minimum distances theoretically (by Kepler's laws) on 21 June/21 December, in reality, on 5 July/5 January in 2020 shifting to 15 July/15 January in 2600. Therefore, the Northern hemisphere should get the extra solar heating in the first 6.5 months because of the shift of a longer distance (local aphelion) to 15 July, which is not fully compensated later by its expected cooling because of the shift of the shorter distance (local perihelion) to 15 January.

## 7. Changes in terrestrial temperature

This level of the solar irradiance increase because of SIM shown in Figure 10 by the amount of about 20-25 W/m² started in 1700 and is expected to last at least until 2500. This indicates that in the current millennium 1600-2600 there is clear extra deposition of solar radiation into the Earth atmosphere. and it's conversion into the terrestrial atmosphere temperature is a complex process involving exchanges between deposited solar radiation, ocean and atmospheric radiative transfer (Harde, 2017). In fact, Harde (2017) using Schwarzschild-Shuster-type models for radiative transfer of UV solar radiation by atmospheric molecules including CO2 have shown that even a smaller increase of solar radiation by 5 W/m² leads to a noticeable (60%) part of the terrestrial temperature increase defined by the Sun (60%) and only 40% defined by the $CO_2$ emission. The further increase of solar irradiance derived above from the Sun-Earth distances using the ephemeris would definitely lead to further contribution of the Sun into the observed terrestrial temperature growth. This requires further investigation using the radiative model simulations.

In the current study we can only roughly estimate possible variations of terrestrial temperature based on the observed curves like that by Akasofu, 2010 (Figure 11). This plot shows a clear increase of the baseline terrestrial temperature, or recovery from 'little ice age' after Maunder minimum, with a rate of 0.5C per century (Figure 11). Since the TSI increase by up to 25 W/m² can be expected until 2500 as evaluated above, the increase in the baseline terrestrial temperature from 1700 can be expected by about 4.0 °C in 2500, or by 1.6°C in 2100 and by 1.4 °C in 2020. However, only the further investigation of conversion of solar radiation into terrestrial atmosphere molecules using the radiative model simulations can provide more accurate numbers.

**Figure 11.** Terrestrial temperature variations derived from terrestrial biomass. Source: Akasofu 2010.

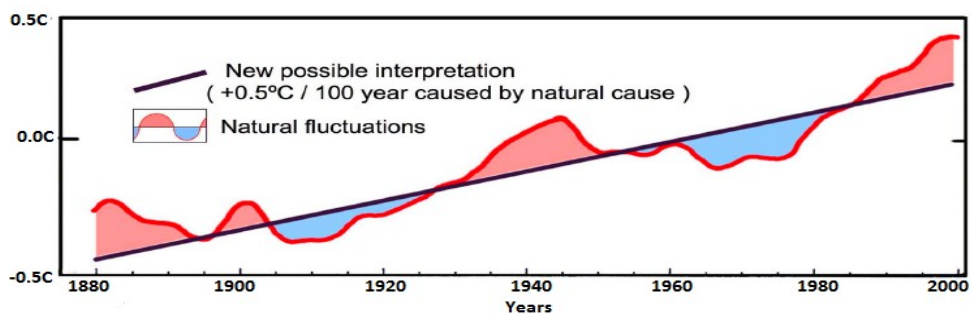

## 8.  Effects of upcoming Grand Solar Minimum (2020-2053)

While a long-term trend of terrestrial temperature variations towards 2500 is warming, in the next 33 years the Sun is entering a period of reduced solar activity, the modern grand solar minimum number 1 (GSM1) occurring in 2020–2053 (Figure 10.5), which can be called a 'mini ice age' as it is twice shorter than Maunder Minimum. There will be also a second modern GSM2 occurring in 2370–2415. The GSMs are caused by the significantly reduced solar magnetic field imposed by the interference of two magnetic s generated by double dynamo in the solar interior (Zharkova et al. 2015).

During the previous GSM (Maunder minimum) solar irradiance was reduced by about 0.22% (Lean et al. 1995; Miller et al. 2012), that, in turn, led to a decrease in the average terrestrial temperature in the Northern Hemisphere of about 1.0 °C as shown in Figure 10.2 (bottom plot) (Lamb 1972; Shindell et al, 2001; Miller et al. 2012; Easterbrook 2016). During these modern GSMs the similar decrease about 0.22% of solar irradiance is expected. Even with the solar irradiance and terrestrial temperature increase caused by the SIM effects discussed above, the terrestrial temperature during this first modern GSM1 is expected to drop by about 1.0C to become only about (1.4-1.0) 0.4 °C higher than that in 1700. The temperature decrease during the second modern GSM (2375–2415) is expected to be about 1.9 C higher than in 1700 that is calculated as follows. The current temperature increase is by 1.4C in 2020, which should increase by 2375 by about another 1.5C (=3 x 0.5C) giving the total increase since 1700 by 2.9 C. The temperature decrease caused by a reduction of solar magnetic field and solar activity during a GSM leads to reduction of temperature by about 1.0 C that results in the total temperature during the GSM2 to be 1.9C higher than in 1700.

After the modern GSMs, solar activity is expected to return to normal (Zharkova et al. 2015) while the terrestrial temperature should following the solar irradiance variations with the effects of SIM, the eleven-year sunspot cycles and terrestrial processes of the conversion of the solar irradiance by the terrestrial atmosphere molecules and ocean into radiation and atmospheric heating.

## 9. Conclusions

Changes in the total solar irradiance reaching the Earth atmosphere depend on regular changes of solar activity in the eleven-year sunspot cycles, in the grand solar cycles generated by a double dynamo and the changes in the Sun-Earth distances caused by orbital perturbations caused by the gravitational forces of the large planets, or solar inertial motion.

TSI reductions caused by a decrease of magnetic field (and solar activity) during GSMs occurring every 350 to 400 years are about 3 W/m², or 0.22%. The TSI increases caused by changes of the Sun-Earth distances since 1600 imposed by SIM in the current Hallstatt cycle (1600 –2600 AD) reached 20 W/m² (1.7%) in 2020, and is further expected to reach 25 W/m² (1.8%) in 2400-2500. These orbital TSI variations definitely exceed the TSI variations caused by the eleven-year sunspot cycles, which approach about 1.3 W/m², or 0.1% of TSI. Hence, in the current millennium M2 (1600-2600) a long trend is that the solar irradiance will keep increasing owing to the orbital SIM effects until about 2500. This would mean that a long-term trend (to year 2500) in the terrestrial temperatures can be added warming from this extra solar input separate from any other reasons.

However, in this year of 2020 the Sun has entered a period of reduced solar activity: the Grand Solar Minimum (2020-2053). This quiet Sun is caused by a significantly reduced magnetic field, which is generated by the interference of double dynamo magnetic waves (Zharkova et al. 2015). This means that during the GSM solar irradiance will be reduced by about 3 W/m², or 0.22%. Therefore, the reduction of solar irradiance caused by the GSM effect will work in opposition to the increase of solar irradiance caused by the orbital SIM effects.

For example, in millennium M2, because of the SIM effects, the baseline temperature (not including any terrestrial effects) is increased by 1.4C since 1700, while during the modern GSM1 it is expected to be lowered by 1.0 C giving the resulting temperature in 2020-2053 being only 0.4 °C higher than in 1700. After 2053, the solar irradiance and the baseline terrestrial temperature is expected to return to the pre-GSM level, then the irradiance and temperature will continue increasing because of the SIM effects combined with radiative transfer of solar radiation in the terrestrial atmosphere. This means the terrestrial temperature will continue increasing up to 2.9-3.0 C until the second modern GSM2 (2375–2415), during which the temperature can be expected to reduce again by 1.0 C to reaching magnitudes which are higher by 1.9-2.0 C than in 1700.

Based on our analysis, the usage of the principal components of the solar background magnetic field as a new and more accurate proxy of solar activity has opened new perspectives for reliable prediction of solar activity on short, medium and long-terms. This approach has allowed us to link these magnetic field variations to the variations of solar irradiance, which are associated with the inner solar processes and with the orbital effects on the Sun-Earth distances. The fundamental oscillations of solar irradiance, in turn, may be linked to the oscillations of the baseline terrestrial temperature, independent of any terrestrial processes of radiative transfer and heating.

**Acknowledgement.** The author V. Zharkova wishes to express her deepest gratitude to the funding by the public supporters raised through Fund-me sites. Their support inspired the author to undertake the investigation of the ephemeris of the Sun-Earth distances and relevant variations of the solar irradiance associated with the changes of the Sun-Earth distances induced by orbital effects.